\documentclass[12pt]{article}
\usepackage{amsmath,amssymb,amscd,amsbsy,amsgen,amsopn,amstext,
amsxtra}

\usepackage[dvips]{graphicx}

\begin{document}

\begin{flushright}
%May 15, 2013
\end{flushright}

\vskip 0.5 truecm

\begin{center}
{\Large{\bf Conditionally valid uncertainty relations}}
\end{center}
\vskip .5 truecm
\centerline{\bf  Kazuo Fujikawa 
}
\vskip .4 truecm
\centerline {\it %$^1$ 
Mathematical Physics Laboratory, RIKEN 
Nishina Center,}
\centerline {\it Wako 351-0198, 
Japan}

\vskip 0.5 truecm

\makeatletter
%\@addtoreset{equation}{section}
%\def\theequation{\thesection.\arabic{equation}}
\makeatother

\begin{abstract}
It is shown that the well-defined unbiased measurement or disturbance of a dynamical variable is not maintained for the precise measurement of the conjugate variable, independently of uncertainty relations. The conditionally valid uncertainty relations on the basis of those additional assumptions, which include most of the familiar Heisenberg-type relations, thus become singular for the precise measurement. We  clarify some contradicting conclusions in the literature concerning those conditionally valid uncertainty relations: The
failure of a naive Heisenberg-type error-disturbance relation and the modified Arthurs-Kelly relation in the recent spin measurement is attributed to this singular behavior. 
    The naive Heisenberg-type error-disturbance relation is formally preserved  in quantum estimation theory, which is shown to be based on the strict unbiased measurement and disturbance, but it leads to unbounded disturbance for bounded operators such as spin variables. In contrast,
the Heisenberg-type error-error uncertainty relation and the Arthurs-Kelly relation, as conditionally valid uncertainty relations, are consistently maintained. 
\end{abstract}

\section{Introduction}   
The uncertainty relation~\cite{heisenberg} has a long history and its essential aspects are well-described by the formulations of Kennard~\cite{kennard} and Robertson~\cite{robertson}.
A recent experiment~\cite{hasegawa}, which invalidated a naive
Heisenberg-type error-disturbance relation~\cite{ozawa1},
revived our interest in this old subject. In contrast to the naive Heisenberg-type error-disturbance relation, the relations which are based on only the positive definite Hilbert space and natural commutator algebra are expected to be valid as long as quantum mechanics is valid, namely, "universally valid"~\cite{ozawa1, ozawa2, fujikawa, fujikawa3}. It was recently shown~\cite{fujikawa3} that all the known universally valid uncertainty relations are derived from Robertson's relation written for suitable combinations of operators. It is important to distinguish 
the uncertainty relations which are universally valid from those relations based on additional assumptions and thus only conditionally valid.  

In this paper, we analyze the implications of the assumptions of unbiased joint measurements or unbiased measurement and disturbance which are widely used in the formulation of uncertainty relations~\cite{arthurs, arthurs2, appleby}. We clarify the origin of quite different conclusions concerning the conditionally valid Heisenberg-type relations in the measurement operator formalism~\cite{ozawa1, ozawa2} and in the quantum estimation theory~\cite{watanabe1, watanabe2} which is a new approach to uncertainty relations. It is first pointed out that  the well-defined unbiased measurement or disturbance of a quantum mechanical  operator is not maintained for the precise measurement of the conjugate operator in the framework of the ordinary measurement theory~\cite{ozawa2, appleby, neumann, kraus, nielsen} .  The conditionally valid uncertainty relation such as the naive Heisenberg-type error-disturbance relation~\cite{hasegawa, ozawa1},  which is  based on the assumptions of unbiased measurement and disturbance, thus fails if one formulates the relation in terms of well-defined bounded operators. We next point out that the consistent estimator in quantum estimation theory~\cite{watanabe1,watanabe2} is equivalent to the assumtions of unbiased measurement and disturbance. The naive Heisenberg-type error-disturbance relation is formally preserved in quantum estimation theory,  but the disturbance of the bounded operator is forced to be singular and divergent for the precise measurement of the conjugate variable\cite{watanabe1,watanabe2}. In contrast, the Heisenberg-type error-error uncertainty relation and the Arthurs-Kelly (and Arthurs-Goodman) relation, as conditionally valid uncertainty relations, are consistently maintained.  The implications of this
analysis on the experimental tests of various forms of uncertainty relations are discussed.   

\section{Conditionally valid uncertainty relations}

\subsection{Algebraic inconsistency}

For the measurement operators of $M$ and $N$~\cite{ozawa2,appleby, neumann, kraus, nielsen} for non-commuting conjugate variables $A$ and $B$, respectively,
the unbiased measurement implies the relations
\begin{eqnarray}
\langle M^{out}- A\rangle=0, \ \ \ \ \langle N^{out}- B\rangle=0
\end{eqnarray}
for {\em all} the states $\psi$ in $|\psi\otimes \xi\rangle$; $|\psi\rangle$ and 
$|\xi\rangle$ stand for the states of the system and the apparatus, respectively. In general, the dimensionality of $|\xi\rangle$ is much larger than that of $|\psi\rangle$. 

We work in the Heisenberg picture and the variables without any suffix stand for the initial variables, and the variables $M^{out}=U^{\dagger}(1\otimes M)U$ and $N^{out}=U^{\dagger}(1\otimes N)U$, for example, stand for the variables after the measurement. The unitary operator $U=U(t_{\rm final})=\exp\{-\frac{i}{\hbar}\hat{H}t_{\rm final}\}$, which depends on $A$, $B$, $M$ and $N$ among others through the total Hamiltonian $\hat{H}$, generates the time development of operators during the measurement interaction, and we assume $U(t)=U(t_{\rm final})$ for $t>t_{\rm final}$. The main part of the analysis in Section 2 is understood without using the details of measurment theory. See Section 3.1 below for further details of measurement theory. The present setting is convenient to formulate
the Arthurs-Kelly-type relations~\cite{arthurs,arthurs2,braustein,stenholm}. To make the mathematics better defined we deal with bounded operators $A$ and $B$ in the following unless stated otherwise.

On the other hand the precise measurement implies, for example,
\begin{eqnarray}
\langle (M^{out}- A)^{2}\rangle=0 \ \  {\rm or} \ \  (M^{out}- A)|\psi\otimes \xi\rangle=0
\end{eqnarray}
for arbitrary given fixed operator $A$ and fixed state $|\psi\rangle$ by suitably choosing 
the measurement operator $M$ and the state $|\xi\rangle$. The precise measurement in this definition does not necessarily imply the unbiased measurement which is valid for all $\psi$. In our application, it is convenient to consider a specific precise measurement such as a precise projective measurement which is a special case of the unbiased measurement.  
 
 In the ordinary measurement theory~\cite{ozawa2,appleby, neumann, kraus, nielsen} we assume $[M, N]=0$, and thus for the unitary time development we have
\begin{eqnarray}
\langle [M^{out}, N^{out}]\rangle=0.
\end{eqnarray}
This relation when combined with the precise measurement of $A$ implies
\begin{eqnarray}
\langle [M^{out}, N^{out}]\rangle&=&\langle [A, N^{out}]\rangle\nonumber\\
&=&\langle [A, N^{out}-B]\rangle+\langle [A, B]\rangle\nonumber\\
&=&\langle [A, B]\rangle
\end{eqnarray}
where we used the relation valid for the unbiased measurement, 
\begin{eqnarray}
\langle [A, N^{out}-B]\rangle=0
\end{eqnarray}
by noting the following identity shown in Appendix of~\cite{appleby}
\begin{eqnarray}
\langle\psi\otimes\xi|{\cal B}|\psi^{\prime}\otimes\xi\rangle&=&
\frac{1}{4}\{
\langle(\psi+\psi^{\prime})\otimes\xi|{\cal B}|(\psi+\psi^{\prime})\otimes\xi\rangle\nonumber\\
&&-\langle(\psi-\psi^{\prime})\otimes\xi|{\cal B}|(\psi-\psi^{\prime})\otimes\xi\rangle\nonumber\\
&&-i\langle(\psi+i\psi^{\prime})\otimes\xi|{\cal B}|(\psi+i\psi^{\prime})\otimes\xi\rangle\nonumber\\
&&+i\langle(\psi-i\psi^{\prime})\otimes\xi|{\cal B}|(\psi-i\psi^{\prime})\otimes\xi\rangle\}
\end{eqnarray}
with ${\cal B}=N^{out}-B$ and $\psi^{\prime}=A\psi$, for example. Note that we have only the "diagonal" elements on the right-hand side, which satisfy the condition of the unbiased measurement $\langle \psi\otimes\xi|{\cal B}|\psi\otimes\xi\rangle=0$ for {\em all} $\psi$.

We thus conclude
\begin{eqnarray}
\langle [M^{out}, N^{out}]\rangle=\langle [A, B]\rangle
\end{eqnarray}
which is a contradiction since the conjugate variables satisfy $\langle [A, B]\rangle\neq 0$ in general, except for the very special state with $\langle [A, B]\rangle=0$. The assumption of unbiased joint measurements does not lead to any apparent contradictions  and  
the precise measurement of $A$ does not contradict the unbiased measurement of $A$ itself, but if one combines the precise measurement of $A$ with the unbiased measurement of the conjugate variable $B$, one recognizes the clear contradiction. The precise projective measurement is included in the unbiased measurement, and thus the assumption of the unbiased joint measurements is algebraically inconsistent for well-defined operators. 

The relation (7) may be interpreted that the precise measurement of $A$ does not allow the unbiased measurement of $B$, if all the operators involved are assumed to be well-defined, since it forces the state $|\psi\rangle$ to be a very specific state which satisfies $\langle\psi|[A, B]|\psi\rangle=0$ to be consistent with $\langle [M^{out}, N^{out}]\rangle=0$.
This interpretation is consistent with the recent analysis of the error-error uncertainty relation in quantum estimation theory~\cite{watanabe1}. We later explain that the quantum estimation theory  imposes very strict unbiased measurements. The authors in~\cite{watanabe1} show on the basis  of numerical simulations that the unbiased measurement of $B$ for the precise projective measurement of $A$ is maintained by leading to the singular $\sigma(N^{out})^{2}=\infty$ and, as a result, the Heisenberg-type error-error uncertainty relation is also satisfied. This provides a very attractive interpretation of the algebraic inconsistency (7). In fact, by noting
$\langle[M^{out},N^{out}]\rangle=0$, we  have
\begin{eqnarray}
|\langle[A,B]\rangle|&=&|\langle[M^{out}-A,N^{out}]\rangle+\langle[A,N^{out}-B]\rangle|\nonumber\\
&\leq& 2||(M^{out}-A)|\psi\otimes\xi\rangle||||N^{out}|\psi\otimes\xi\rangle||\nonumber\\
&\leq& 2||(M^{out}-A)|\psi\otimes\xi\rangle||||N^{out}||
\end{eqnarray}
if one assumes the unbiased measurement $\langle[A,N^{out}-B]\rangle=0$. This relation shows that $||N^{out}||\rightarrow\infty$ for $||(M^{out}-A)|\psi\otimes\xi\rangle||\rightarrow 0$ is consistent with $|\langle[A,B]\rangle|\neq 0$. 
We discuss this issue in more detail later.

Similarly, one concludes 
\begin{eqnarray}
\langle [M^{out}, B^{out}]\rangle=\langle [A, B]\rangle
\end{eqnarray}
if one assumes the precise measurement of $A$ and the unbiased disturbance of 
$B$ which implies $\langle B^{out} - B \rangle=0$ for all $\psi$. Here $B^{out}=U^{\dagger}(B\otimes 1)U$ 
stands for the variable $B$ after the measurement of $A$. This relation contradicts the ordinary assumption of the independence of the dynamical
variable and measurement apparatus specified by $[M, B]=0$.  The precise projective measurement is included in the unbiased measurement, and thus the assumption of the unbiased measurement and disturbance is algebraically inconsistent for well-defined operators.

Here again, we interpret the algebraic inconsistency (9) as an indication of the absence of the unbiased disturbance of $B$ for the precise projective measurement of $A$, if all the operators involved are well-defined. In the present case, however, we prefer to keep the disturbance of the {\it bounded} operator finite $\sigma(B^{out})\leq ||B||$ to be consistent with the conventional notion of disturbance instead of forcing $\sigma(B^{out})$ to be unbounded and singular, and thus the unbiased disturbance condition is forced to fail. By noting $\langle [M^{out}, B^{out}]\rangle=0$, we have
\begin{eqnarray}
|\langle[A,B]\rangle|&=&|\langle[M^{out}-A,B^{out}]\rangle+\langle[A,B^{out}-B]\rangle|\\
&\leq& 2||(M^{out}-A)|\psi\otimes\xi\rangle||||B^{out}||+|\langle[A,B^{out}-B]\rangle|\nonumber
\end{eqnarray}
which is consistent with $||B^{out}||=||B||$ and $|\langle[A,B]\rangle|\neq 0$
for $||(M^{out}-A)|\psi\otimes\xi\rangle||\rightarrow 0$ if one has 
$\langle [A, B^{out}]\rangle\rightarrow 0$ that is natural for the precise projective measurement of $A$, for which $B^{out}=\sum_{k}P_{k}BP_{k}$ with the spectral decomposition $A=\sum_{k}a_{k}P_{k}$. See also Section 3.1 below for the notational details of measurement theory. In any case, the unbiased disturbance condition 
\begin{eqnarray}
\langle[A,B^{out}-B]\rangle=0
\end{eqnarray}
 for all $\psi$ inevitably fails for $|\langle[A,B]\rangle|\neq 0$. 

Because of the continuity argument, the Heisenberg-type error-disturbance relation, which is based on the unbiased measurement and disturbance, then fails in the broader range of measurement processes not restricted to the precise measurement of $A$. To be more precise, the derivation of the naive Heisenberg-type error-disturbance relation from Robertson's relation fails because of the failure of the assumption of unbiased disturbance~\cite{fujikawa3}. In any case, the operational definition of unbiased disturbance as such is ill-defined since we have no control of the distribution of $B$ after the measurement of the conjugate variable $A$, and thus it may be natural to accept the failure of the naive Heisenberg-type error-disturbance relation as a result of the failure of the unbiased disturbance.  We discuss this issue in more detail later in connection with the analysis of the error-disturbance uncertainty relation.

The analysis in this subsection shows that the condition of unbiased joint measurements or unbiased measurement
and disturbance gives rise to a constrained system in the analysis of algebraic properties of linear operators in quantum mechanics, and thus it is not surprising if the behavior unfamiliar in the conventional representation theory of linear operators appears. 

\subsection{Consistency of the Arthurs-Kelly relation}

To illustrate the implications of our analysis in the preceding subsection, we discuss the Arthurs-Kelly uncertainty relation and related uncertainty relations.

We start with Robertson's relation~\cite{robertson}
\begin{eqnarray}
\sigma(M^{out}-A)\sigma(N^{out}-B)\geq\frac{1}{2}|\langle [M^{out}-A, 
N^{out}-B]\rangle| 
\end{eqnarray}
which is valid for any hermitian operators $M^{out}-A$ and $N^{out}-B$ and thus truly universal. The right-hand side of this relation is written using the triangle inequality (in the two-dimensional space of complex numbers) in the form  
\begin{eqnarray}
&&\sigma(M^{out}-A)\sigma(N^{out}-B)\nonumber\\
&&\geq \frac{1}{2}|\langle-[M^{out},N^{out}]+ [M^{out}, N^{out}-B]+[M^{out}-A,N^{out}]\nonumber\\
&&\hspace{0.8cm} +[-A,-B]\rangle|\nonumber\\
&&\geq \frac{1}{2}\{|\langle [A,B]\rangle|-|\langle [M^{out},N^{out}-B]\rangle|-|\langle [M^{out}-A,N^{out}]\rangle|\nonumber\\
&&\hspace{0.8cm} -|\langle [M^{out},N^{out}]\rangle|\},
\end{eqnarray}
from which one obtains a universally valid relation (without assuming $[M^{out},N^{out}]=0$ in general) using the suitable variations of  Robertson's relation (12) such as $\sigma(M^{out}-A)\sigma(N^{out})\geq \frac{1}{2}|\langle [M^{out}-A,N^{out}]\rangle|$,
\begin{eqnarray}
\{\sigma(M^{out}-A)+\sigma(M^{out})\}\{\sigma(N^{out}-B)+\sigma(N^{out})\}
\geq  \frac{1}{2}|\langle [A,B]\rangle|,
\end{eqnarray}
namely, a Heisenberg-type relation. In this relation  we do not make any extra assumptions such as the unbiased joint measurements. This relation is amusing since it holds for any hermitian $M^{out}$ and $N^{out}$.

If one makes the ordinary assumption $[M^{out},N^{out}]=0$, Robertson's relation is re-written  using the triangle inequality as 
\begin{eqnarray}
&&\sigma(M^{out}-A)\sigma(N^{out}-B)\nonumber\\
&&\geq \frac{1}{2}|\langle [-A,N^{out}-B]+[M^{out}-A,-B]-[-A,-B]\rangle|\nonumber\\
&&\geq \frac{1}{2}\{|\langle [A,B]\rangle|-|\langle [A,N^{out}-B]\rangle|-|\langle [M^{out}-A,B]\rangle|\}.
\end{eqnarray}
From this relation combined with the suitable variations of  Robertson's relation  such as $\sigma(M^{out}-A)\sigma(B)\geq \frac{1}{2}|\langle [M^{out}-A,B]\rangle|$, one can derive a universally valid relation
\begin{eqnarray}
\{\sigma(M^{out}-A)+\sigma(A)\}\{\sigma(N^{out}-B)+\sigma(B)\}
\geq |\langle [A,B]\rangle|,
\end{eqnarray}
while one obtains directly from (15)
\begin{eqnarray}
\sigma(M^{out}-A)\sigma(N^{out}-B)
\geq \frac{1}{2}|\langle [A,B]\rangle|,
\end{eqnarray}
if one assumes the unbiased joint measurements using (5) and (6). 
We now note 
\begin{eqnarray}
&&\epsilon(A)\equiv \langle (M^{out}-A)^{2}\rangle
\geq \sigma(M^{out}-A),\nonumber\\
&&\epsilon(B)\equiv \langle (N^{out}-B)^{2}\rangle
\geq \sigma(N^{out}-N),
\end{eqnarray}
and the relation (17), for example, is written as a Heisenberg-type error-error relation
\begin{eqnarray}
\epsilon(A)\epsilon(B)
\geq \frac{1}{2}|\langle [A,B]\rangle|.
\end{eqnarray}
The quantity 
$\sigma(M^{out}-A)$ in Robertson's relation (12), for example,  is originally defined as an average
of the operator $M^{out}-A$ using the state $|\psi\otimes\xi\rangle$ for any given $M^{out}$, in principle independently of the joint measurements of $A$ and $B$. But after the above replacement, the quantity $\epsilon(A)$ is interpreted as an "error" in the joint  measurements of $A$ and $B$ by assigning  specific time development to $M^{out}$ in the Heisenberg picture~\cite{ozawa1}~\footnote{It would be interesting to work in the Schr\"{o}dinger picture which emphasizes different aspects~\cite{distler}.}.

One can derive the standard Arthurs-Kelly relation from the uncertainty relation for the joint measurements (19) and the simplest Robertson's relation, $\sigma(A)\sigma(B)\geq \frac{1}{2}|\langle [A,B]\rangle|$, as~\cite{appleby}
\begin{eqnarray}
&&\{\epsilon(A)^{2}+ \sigma(A)^{2}\}\{\epsilon(B)^{2}+ \sigma(B)^{2}\}\nonumber\\
&&\geq \frac{1}{4}|\langle [A,B]\rangle|^{2}\{\epsilon(B)^{-2}+ \sigma(B)^{-2}\}\{\epsilon(B)^{2}+ \sigma(B)^{2}\}\nonumber\\
&&\geq |\langle [A,B]\rangle|^{2}.
\end{eqnarray}
Note that the relation (19) is more accurate than (20) as an inequality.
One may combine the relation (20) with
\begin{eqnarray}
\langle M^{out}\rangle&=&\langle M^{out}-A\rangle + \langle A\rangle, \nonumber\\
&=&\langle A\rangle,\nonumber\\
\langle (M^{out})^{2}\rangle&=&\langle (M^{out}-A)^{2}\rangle + \langle A^{2}\rangle+\langle (M^{out}-A)A\rangle+\langle A(M^{out}-A)\rangle\nonumber\\
&=&\langle (M^{out}-A)^{2}\rangle + \langle A^{2}\rangle
\end{eqnarray}
where we assumed the unbiased measurement, and thus 
\begin{eqnarray}
\sigma(M^{out})^{2}=\epsilon(A)^{2}+\sigma(A)^{2}
\end{eqnarray}
and similarly for $\sigma(N^{out})^{2}$. We thus obtain the standard Arthurs-Kelly (and Arthurs-Goodman) relation~\cite{arthurs, arthurs2}
\begin{eqnarray}
\sigma(M^{out})\sigma(N^{out})\geq |\langle [A,B]\rangle|.
\end{eqnarray}
The validity of this relation has been analyzed in the past~\cite{she, yuen,yamamoto}. 

On the other hand, the universally valid Arthurs-Kelly relation~\cite{fujikawa}, which is  derived from Robertson's relation (12) and (16), is written as 
\begin{eqnarray}
\bar{\epsilon}(A)\bar{\epsilon}(B)\geq |\langle [A,B]\rangle|
\end{eqnarray}
where
\begin{eqnarray}
\bar{\epsilon}(A)&\equiv&\epsilon(A)+\sigma(A)\nonumber\\
&=&\langle (M^{out}-A)^{2}\rangle^{1/2}+\langle (A-\langle A\rangle)^{2}\rangle^{1/2},\nonumber\\
\bar{\epsilon}(B)&\equiv&\epsilon(B)+\sigma(B)\nonumber\\
&=&\langle (N^{out}-B)^{2}\rangle^{1/2}+\langle (B-\langle B\rangle)^{2}\rangle^{1/2}.
\end{eqnarray}
Here we assume $\langle [M^{out}, N^{out}]\rangle=0$ but do not assume  the unbiased joint measurements. The saturation of Robertson's relation (12) is a {\em necessary condition} of the saturation of the universally valid Arthurs-Kelly relation (24), and the direct evaluation of the right-hand side of the second line in (15) 
\begin{eqnarray}
\epsilon(A)\epsilon(B)%\nonumber\\
\geq \frac{1}{2}|\langle [A,N^{out}-B]\rangle+\langle [M^{out}-A,B]\rangle+\langle [A,B]\rangle|
\end{eqnarray}
is more accurate than (24) as an inequality.

The Heisenberg-type uncertainty relation for the joint measurements in (19), which is derived from the universally valid Robertson's relation (15) or (26) by assuming the unbiased joint measurements, does  not hold for $\epsilon(A)=0$ and  $\frac{1}{2}\langle [A,B]\rangle\neq 0$ if 
$\epsilon(B)^{2}= \langle (N^{out}-B)^{2}\rangle= \langle (N^{out})^{2}\rangle
-\langle B^{2}\rangle$ is well-defined and finite. This is also consistent with the algebraic inconsistency (7). The numerical analysis in quantum estimation theory~\cite{watanabe1} suggests that 
the strict unbiased joint measurements ensure the error-error uncertainty relation  (19) by driving $\epsilon(B)$ to be singular for  $\epsilon(A)=0$ and $\frac{1}{2}\langle [A,B]\rangle\neq 0$, namely, $\epsilon(B)\rightarrow \infty$ for $\epsilon(A)\rightarrow 0$ while maintaining the precise unbiased joint measurements. In this case both of (19) and the standard Arthurs-Kelly relation hold, although those two relations are conditionally valid based on strict unbiased conditions. The uncertainty relation (19) is thus more of the manifestation of the properties of measuring apparatus than the physical system itself.

In contrast, the precise measurement $\epsilon(A)=0$ with well-defined  $\epsilon(B)$ is consistent with the universally valid relation (26), which is equivalent to Robertson's relation 
(12), since the right-hand side of (26) vanishes for  $(M^{out}- A)|\psi\otimes \xi\rangle=0$ combined with $[M^{out}, N^{out}]=0$ for a general state $|\psi\rangle$.  Also, the universally valid version of the Arthurs-Kelly relation (24) holds for the vanishing "inaccuracy" $\bar{\epsilon}(A)=\epsilon(A)+\sigma(A)=0$, namely, $\epsilon(A)=0$ and $\sigma(A)=0$ even for $\bar{\epsilon}(B)<\infty$, since $\sigma(A)=0$ constrains the state $|\psi\rangle$ to be an eigenstate of the discrete eigenvalue of $A$ and thus $\langle [A,B]\rangle=0$. In general, $\epsilon(A)=0$  specifies the measurement apparatus and procedure while 
$\sigma(A)=0$ constrains the dynamical variable $A$ and the physical state if one does not impose any extra conditions.

\subsection{Consistency of error-disturbance uncertainty relations}
 
We next analyze the error-disturbance uncertainty relations~\cite{braginsky, appleby}  which are interesting in view of the recent spin measurement.
 If one chooses
\begin{eqnarray}
A=\sigma_{x},\ \ \ \ \, B=\sigma_{y},\ \ \ \ \, \psi=|+ z\rangle,
\end{eqnarray}
one reproduces the set-up of the spin measurement in~\cite{hasegawa}.
We examine the consistency of error-disturbance uncertainty relations in view of the actual experimental set up of the spin measurement later.

We start with Robertson's relation
\begin{eqnarray}
\sigma(M^{out}-A)\sigma(B^{out}-B)\geq \frac{1}{2}|\langle [M^{out}-A,B^{out}-B]\rangle|
\end{eqnarray}
and the actual analysis proceeds parallel to the analysis in the preceding subsection. In fact the analysis of this problem was presented in~\cite{fujikawa3} and we here briefly summarize the main results.
The most general Heisenberg-type relation
\begin{eqnarray}
\{\sigma(M^{out}-A)+\sigma(M^{out})\}\{\sigma(B^{out}-B)+\sigma(B^{out})\}
\geq  \frac{1}{2}|\langle [A,B]\rangle|
\end{eqnarray}
is derived from (28) without assuming $\langle [M^{out}, B^{out}]\rangle=0$. 
This relation is valid for {\em any} hermitian $M^{out}$ and $B^{out}$.

By assuming $[M^{out},B^{out}]=0$ and 
using the triangle inequality (in the two-dimensional space of complex numbers) one can derive the following relation from Robertson's relation (28), 
\begin{eqnarray}
&&\sigma(M^{out}-A)\sigma(B^{out}-B)\nonumber\\
&&\geq \frac{1}{2}|\langle [-A,B^{out}-B]+[M^{out}-A,-B]-[-A,-B]\rangle|\nonumber\\
&&\geq \frac{1}{2}\{|\langle [A,B]\rangle|-|\langle [A,B^{out}-B]\rangle|-|\langle [M^{out}-A,B]\rangle|\}.
\end{eqnarray}
Using the suitable variations of  Robertson's relation  such as $\sigma(M^{out}-A)\sigma(B)\geq \frac{1}{2}|\langle [M^{out}-A,B]\rangle|$, one derives   from (30) the universally valid relation,
\begin{eqnarray}
\{\sigma(M^{out}-A)+\sigma(A)\}\{\sigma(B^{out}-B)+\sigma(B)\}
\geq |\langle [A,B]\rangle|,
\end{eqnarray}
while one obtains directly from (30)
\begin{eqnarray}
\sigma(M^{out}-A)\sigma(B^{out}-B)
\geq \frac{1}{2}|\langle [A,B]\rangle|,
\end{eqnarray}
if one assumes the unbiased measurement and disturbance which imply
$\langle [M^{out}-A,B]\rangle=0$ and $\langle [A,B^{out}-B]\rangle=0$.

Eq.(31) leads to a relation, which was suggested to be called 
"universally valid Heisenberg relation" in~\cite{fujikawa}~\footnote{In the present paper we 
analyze only the "Heisenberg-type" uncertainty relations where a product of two factors referring to conjugate variables appears on the left-hand side. The relation proposed by Ozawa~\cite{ozawa1} consists of a sum of three terms on the left-hand side; $\epsilon(A)\eta(B)+\sigma(A)\eta(B)+\epsilon(A)\sigma(B)\geq \frac{1}{2}|\langle [A,B]\rangle|$.},  
\begin{eqnarray}
\bar{\epsilon}(A)\bar{\eta}(B)\geq |\langle [A,B]\rangle|
\end{eqnarray}
where
\begin{eqnarray}
\bar{\epsilon}(A)&\equiv&\epsilon(A)+\sigma(A)\nonumber\\
&=&\langle (M^{out}-A)^{2}\rangle^{1/2}+\langle (A-\langle A\rangle)^{2}\rangle^{1/2},\nonumber\\
\bar{\eta}(B)&\equiv&\eta(B)+\sigma(B)\nonumber\\
&=&\langle (B^{out}-B)^{2}\rangle^{1/2}+\langle (B-\langle B\rangle)^{2}\rangle^{1/2}.
\end{eqnarray}
In the derivation of (33) from (31), we used the relation 
\begin{eqnarray}
\eta(B)\equiv \langle (B^{out}-B)^{2}\rangle \geq \sigma(B^{out}-B),
\end{eqnarray}
where $\eta(B)$ stands for the "disturbance", in addition to the "error"
$\epsilon(A)$ in (18).
Here again, the quantity 
$\sigma(B^{out}-B)$ in Robertson's relation (28), for example, is originally defined as an average
of the operator $B^{out}-B$ using the state $|\psi\otimes\xi\rangle$ for any given $B^{out}$, in principle independently of the measurement of $A$. But after the above replacement, the quantity $\eta(B)$ is interpreted as a disturbance caused by the measurement of $A$ by assigning  specific time development to $B^{out}$ in the Heisenberg picture~\cite{ozawa1}.
The saturation of Robertson's relation (28) is a {\em necessary condition} of the saturation of the universally valid Heisenberg relation (33), and the direct evaluation of the right-hand side of the second line in Robertson's relation (30) 
\begin{eqnarray}
\epsilon(A)\eta(B)%\nonumber\\
\geq \frac{1}{2}|\langle [A,B^{out}-B]\rangle+\langle [M^{out}-A,B]\rangle+\langle [A,B]\rangle|
\end{eqnarray}
is more accurate than (33) and also the version proposed by Ozawa~\cite{ozawa1}, both of which are the secondary consequences of the universally valid Robertson's relation (36)~\cite{fujikawa3}. The evaluation of the right-hand side of (36) is performed using the identity (6), for example, following  essentially the same steps as the evaluation of $\epsilon(A)$ and $\eta(B)$.

One can derive the naive Heisenberg-type error-disturbance relation~\cite{ozawa1} 
\begin{eqnarray}
\epsilon(A)\eta(B)\geq \frac{1}{2}|\langle [A,B]\rangle|
\end{eqnarray}
from (32) which is based on the unbiased measurement and disturbance, as was emphasized in~\cite{fujikawa3}. Algebraically the derivation of this relation is not justified for the precise measurement $\epsilon(A)=0$ with $\frac{1}{2}|\langle [A,B]\rangle|\neq 0$ if all the operators involved are assumed to be well-defined and thus for a bounded operator $B$ with $\eta(B)<\infty$, as was shown in (9). Actually, the left-hand side of this relation vanishes for the precise measurement of $A$ independently of the value of the right-hand 
side $\frac{1}{2}|\langle [A,B]\rangle|$ for the well-defined  $\eta(B)$, which is generally bounded by the norm $||B^{out}-B||$ for the bounded operator~\cite{ozawa2}. For unbounded operators, this argument is technically subtle but the algebraic inconsistency we discussed is expected to persist if properly formulated. The relation (37) was invalidated by the recent spin measurement~\cite{hasegawa}. The relation (37) in the context of quantum estimation theory will be discussed in the next section.  

By combining the naive error-disturbance relation (37) with the simplest Robertson's relation 
$\sigma(A)\sigma(B)\geq \frac{1}{2}|\langle [A,B]\rangle|$,
one obtains
\begin{eqnarray}
&&\{\epsilon(A)^{2}+\sigma(A)^{2}\}\{\eta(B)^{2}+\sigma(B)^{2}\}
\nonumber\\
&&\geq \frac{1}{4}|\langle [A,B]\rangle|^{2}\{\eta(B)^{-2}+\sigma(B)^{-2}\}\{\eta(B)^{2}+\sigma(B)^{2}\}\nonumber\\
&&\geq |\langle [A,B]\rangle|^{2}.
\end{eqnarray}
If one uses the relations 
\begin{eqnarray}
&&\sigma(M^{out})^{2}=\epsilon(A)^{2}+\sigma(A)^{2},\nonumber\\
&&\sigma(B^{out})^{2}=\eta(B)^{2}+\sigma(B)^{2},
\end{eqnarray}
which hold if one assumes the unbiased measurement and disturbance such as 
$\langle (M^{out}-A)A\rangle=0$ and $\langle (B^{out}-B)B\rangle=0$, one obtains from (38) the {\em modified} Arthurs-Kelly relation ( in contrast to the standard Arthurs-Kelly relation (23) which contains $\sigma(M^{out})\sigma(N^{out})$)~\cite{fujikawa}
\begin{eqnarray}
\sigma(M^{out})\sigma(B^{out})\geq |\langle [A,B]\rangle|.
\end{eqnarray}
This relation is closely related to the universally valid Heisenberg uncertainty relation (33), but it is shown~\cite{fujikawa3} that the relation (33) always holds while the relation (38) fails for the spin-measurement~\cite{hasegawa}. This failure of (38) is expected, though not proved, from the failure of the naive error-disturbance relation (37).

\subsection{Spin measurement experiment}
We now comment on the algebraic consistency of error-disturbance
uncertainty relations  in view of the actual spin measurement~\cite{hasegawa}. 
Their experiment is based on the projective measurement defined by
\begin{eqnarray}
\langle M^{out}\rangle&=& \langle\psi|(+1) E_{\phi}(+)+(-1)E_{\phi}(-)|\psi\rangle\nonumber\\
&=&\langle\psi|\sigma_{\phi}|\psi\rangle,\nonumber\\
\langle B^{out}\rangle&=&\langle\psi|E_{\phi}(+)\sigma_{y}E_{\phi}(+)+E_{\phi}(-)\sigma_{y}E_{\phi}(-)]|\psi\rangle\nonumber\\
&=&\langle\psi|\sin\phi\sigma_{\phi}|\psi\rangle
\end{eqnarray}
where $A=\sigma_{x}$, $B=\sigma_{y}$ and the specific eigenstate $\psi=|+z\rangle$ of $\sigma_{z}$. See Section 3.1 below for notational details of measurement theory. The auxiliary operator
\begin{eqnarray}
\sigma_{\phi}=\cos\phi\sigma_{x}+\sin\phi\sigma_{y}
\end{eqnarray}
is introduced with $0\leq \phi\leq\pi/2$ called "detuning" angle~\cite{hasegawa}.
The projection operators are defined by
\begin{eqnarray}
E_{\phi}(\pm )=(1\pm \sigma_{\phi})/2.
\end{eqnarray}
In terms of the parameter $\phi$, $\epsilon(A)=2\sin\frac{\phi}{2}$ in (18),
$\eta(B)=\sqrt{2}\cos\phi$ in (35) and $\sigma(A)=\sigma(B)=1$~\cite{hasegawa}, and  Robertson's relation (36) after evaluating its right-hand side becomes
\begin{eqnarray}
\epsilon(A)\eta(B)=2\sqrt{2}\sin\frac{\phi}{2}\cos\phi\geq 2\sin^{2}\frac{\phi}{2}\cos\phi.
\end{eqnarray}
The difference of both-hand sides $\Delta=2\sin\frac{\phi}{2}\cos\phi(\sqrt{2}-\sin\frac{\phi}{2})\geq 0$ for all the "detuning" angle $0\leq \phi\leq\pi/2$. Note that the inequality (44) is saturated at $\phi=0$ and $\phi=\pi/2$. If one is willing to accept $\epsilon(A)$ in (18) as a measurement "error", then $\epsilon(A)=2\sin\frac{\phi}{2}=0$ for finite $\eta(B)=\sqrt{2}\cos\phi$, which is realized at $\phi=0$, is allowed by Robertson's relation (28) or (36).

The universally valid Heisenberg relation (33) becomes
\begin{eqnarray}
(2\sin\frac{\phi}{2}+1)(\sqrt{2}\cos\phi+1)\geq 2
\end{eqnarray}
and this relation as well as Ozawa's original relation~\cite{ozawa1} is satisfied for all $\phi$, but the equality sign is not achieved in either case. 

In contrast, the naive Heisenberg-type error-disturbance relation (37),
\begin{eqnarray}
\epsilon(A)\eta(B)=2\sqrt{2}\sin\frac{\phi}{2}\cos\phi\geq 1
\end{eqnarray}
is shown to fail for all $\phi$~\cite{hasegawa}. This relation is derived from universally valid Robertson's relation (36) if unbiased measurement and unbiased disturbance conditions are satisfied. The unbiased measurement condition $\langle M^{out}\rangle=\langle\psi|\sigma_{\phi}|\psi\rangle=\langle\psi|\sigma_{x}|\psi\rangle$ of $A$ for all $\psi$ is not satisfied by the measurement (41) for $\phi\neq 0$.
For the precise projective measurement of $A$, which is realized for $\phi=0$, the unbiased measurement condition is satisfied, and the unbiased disturbance condition is given by
\begin{eqnarray}
\langle B^{out}-B\rangle&=&\langle\psi|\sin\phi\sigma_{\phi}-\sigma_{y}|\psi\rangle\nonumber\\
&=&\langle\psi|-\sigma_{y}|\psi\rangle=0
\end{eqnarray}
which is satisfied for the eigenstate of $\sigma_{z}$, $\psi=|+z\rangle$, but not all $\psi$. Namely, the derivation of (46) itself is not justified, and in fact we have already shown in (9) that the unbiased measurement and unbiased disturbance conditions are not consistently implemented for well-defined operators. Further discussion of the naive error-disturbance relation (37) is given in the next section. 

In the spin measurement~\cite{hasegawa}, $\bar{\eta}(B)=\eta(B)+\sigma(B)<\infty$ but yet the universally valid 
Heisenberg relation (33) is valid even for the vanishing "inaccuracy" $\bar{\epsilon}(A)=\epsilon(A)+\sigma(A)=0$, namely, for
$\epsilon(A)=\sigma(A)=0$, since $\sigma(A)=0$ constrains the state $|\psi\rangle$ to be an eigenstate of the discrete eigenvalue of $A=\sigma_{x}$ and thus $|\langle [A,B]\rangle|=0$. On the other hand, the universally valid relation (36), which is equivalent to Robertson's relation (28), holds for $\epsilon(A)=0$ for a general state $|\psi\rangle$ since the right-hand side of (36) also vanishes for  $(M^{out}- A)|\psi\otimes \xi\rangle=0$ combined with $[M^{out}, B^{out}]=0$. Here again, $\epsilon(A)=0$  specifies the measurement apparatus and procedure while 
$\sigma(A)=0$ constrains the dynamical variable $A$ and the physical state if no additional conditions are imposed. 

The universally valid Heisenberg relation (33) implies that the original idea of Heisenberg is realized by a combination of these two properties in the form 
$\bar{\epsilon}(A)=\epsilon(A)+\sigma(A)$ suggesting that $\sigma(A)$ gives an intrinsic "error" even for the precise measurement with $\epsilon(A)=0$.
For bounded operators, $\bar{\epsilon}(A)=0$ implies $|\langle [A,B]\rangle|=0$ with $\bar{\eta}(B)<\infty$. For unbounded operators such as $A=\hat{p}$ and $B=\hat{x}$, $\bar{\epsilon}(A)\rightarrow 0$ implies $\bar{\eta}(B)\rightarrow \infty$ because $\sigma(B)\rightarrow\infty$ for $\sigma(A)\rightarrow 0$~\footnote{For $A=\hat{p}$ and $B=\hat{x}$, one can show $\sigma(p)\sigma(x)\geq \frac{1}{2}|\langle\hat{p}\psi,\hat{x}\psi\rangle-\langle\hat{x}\psi,\hat{p}\psi\rangle|=\frac{1}{2}|1-L|\psi(\frac{L}{2})|^{2}|$ for arbitrary large but finite $L$ with a periodic boundary condition $\psi(\frac{L}{2})=\psi(-\frac{L}{2})$ in box normalization~\cite{fujikawa3}; $\sigma(p)=0$ with a discrete eigenvalue of $A=\hat{p}$ is consistent with finite $\sigma(x)\sim L$ since the right-hand side also vanishes. But if one takes $L=\infty$ first and considers only the normalizable states, for which $L|\psi(\frac{L}{2})|^{2}=0$ for $L\rightarrow\infty$, $\sigma(p)=0$ implies $\sigma(x)=\infty$. }.

\section{Quantum estimation theory}

We now examine the analyses of Watanabe, Sagawa and Ueda~\cite{watanabe1} and Watanabe and Ueda~\cite{watanabe2} in more detail. These works are based on quantum estimation theory which is a new framework to study uncertainty relations, and it is important to understand why those authors arrive at the conclusion (in particular in~\cite{watanabe2}) quite different from that in the more conventional formalism~\cite{hasegawa, ozawa1,ozawa2}. 

\subsection{Measurement operator formalism}
We first summarize the basic aspects of the measurement operator formalism~\cite{ozawa2, neumann, kraus, nielsen}. We define
\begin{eqnarray}
M^{out}=U^{\dagger}(1\otimes M)U, 
\end{eqnarray}
in the Heisenberg picture, and in the corresponding Schr\"{o}dinger picture
\begin{eqnarray}
U|\psi\otimes \xi\rangle \equiv \sum_{k,l}|M_{k,l}\psi\otimes \xi_{k,l}\rangle
\end{eqnarray}
with the orthonormal complete simultaneous eigenstates $|\xi_{k,l}\rangle$ of the hermitian $M$ and $N$; $M|\xi_{k,l}\rangle=m_{k}|\xi_{k,l}\rangle$ and $N|\xi_{k,l}\rangle=n_{l}|\xi_{k,l}\rangle$ since $[M, N]=0$ by assumption. The operator $M_{k,l}$ generally depends on the initial apparatus state $|\xi\rangle$, $M_{k,l}=M_{k,l}(\xi)$. Note that the dimensionality of $M$ and $N$ is very large in general. The operator $U=U(t_{\rm final})=\exp\{-\frac{i}{\hbar}\hat{H}t_{\rm final}\}$ depends on $A$, $B$, $M$ and $N$ among others through the total Hamiltonian $\hat{H}$, and we assume $U(t)=U(t_{\rm final})$ for $t>t_{\rm final}$. Eq.(49) shows that the separable state $|\psi\otimes \xi\rangle$ is converted to an entangled state by the measurement Hamiltonian, and the right-hand side of (49) may be regarded as a purification of the mixed physical state $\rho=\sum_{k,l}M_{k,l}|\psi\rangle\langle\psi|M_{k,l}^{\dagger}$ after the measurement.
The unitarity of $U$ implies
\begin{eqnarray}
\sum_{k,l}\sum_{k^{\prime},l^{\prime}}\langle \xi_{k^{\prime},l^{\prime}}\otimes M_{k^{\prime},l^{\prime}}\psi|M_{k,l}\psi\otimes \xi_{k,l}\rangle=\sum_{k,l}\langle \psi|M^{\dagger}_{k,l}M_{k,l}|\psi\rangle=1
\end{eqnarray}
and thus 
\begin{eqnarray}
\sum_{k,l}M^{\dagger}_{k,l}M_{k,l}=\sum_{k,l}E_{k,l}=1
\end{eqnarray}
and $\{E_{k,l}\}=\{M^{\dagger}_{k,l}M_{k,l}\}$ define the positive operator valued measures with the operators $\{ M_{k,l}\}$ standing for the measurement operators of Kraus-type~\cite{kraus}. In this formulation, we have 
\begin{eqnarray}
\langle M^{out}\rangle&=&\sum_{k,l}m_{k}\langle \psi|M^{\dagger}_{k,l}M_{k,l}|\psi\rangle=\sum_{k,l}m_{k} p_{k,l}(\psi),\nonumber\\
\langle (M^{out})^{2}\rangle&=&\sum_{k,l}m_{k}^{2}\langle \psi|M^{\dagger}_{k,l}M_{k,l}|\psi\rangle=\sum_{k,l}m_{k}^{2}p_{k,l}(\psi),
\end{eqnarray}
with $m_{k}$ standing for the eigenvalues of $M$, and thus 
\begin{eqnarray}
\sigma(M^{out})^{2}&=&\langle (M^{out})^{2}\rangle-\langle M^{out}\rangle^{2}
\nonumber\\
&=&\sum_{k,l}p_{k,l}(\psi)(m_{k}-\sum_{k^{\prime},l^{\prime}}m_{k^{\prime}} p_{k^{\prime},l^{\prime}}(\psi))^{2}.
\end{eqnarray}
By noting $B^{out}=U^{\dagger}(B\otimes 1)U$ we also have
\begin{eqnarray}
\langle B^{out}\rangle&=&\sum_{k,l}\langle \psi|M^{\dagger}_{k,l}BM_{k,l}|\psi\rangle,\nonumber\\
\langle (B^{out})^{2}\rangle&=&\sum_{k,l}\langle \psi|M^{\dagger}_{k,l}B^{2}M_{k,l}|\psi\rangle.
\end{eqnarray}
The {\em unbiased} measurement implies
\begin{eqnarray}
\langle M^{out}\rangle=\sum_{k,l}m_{k}\langle \psi|M^{\dagger}_{k,l}M_{k,l}|\psi\rangle=\langle \psi|A|\psi\rangle
\end{eqnarray}
for all $\psi$. By noting an identity similar to (6), one obtains $A = \sum_{k,l}m_{k}M^{\dagger}_{k,l}M_{k,l}$. 
 
\subsection{Consistent estimator and uncertainty relation} 
 The authors in references~\cite{watanabe1, watanabe2} introduce an estimator $A^{est}$ in terms of measured quantities.
The estimator of $A^{est}$ is a function of $\{n_{i}\}$: $A^{est} = A^{est}(\{n_{i}\})$. The set  $\{n_{i}\}$ consists of integers that satisfy $n_{i}\geq 0$ and $\sum_{i} n_{i} = n$, where $n$ stands for the total number of similarly prepared quantum mechanical samples. Their actual analysis is mainly based on positive operator valued measures. As an explicit example, we thus adopt
\begin{eqnarray}
A^{est}(\{n_{i}\})=\sum_{i}m_{i}\frac{n_{i}}{n}
\end{eqnarray}
corresponding to the positive operator valued measures in (52). The general estimation theory should work for this ideal choice of $A^{est}(\{n_{i}\})$ also.

The expectation value and variance of the estimator $A^{est}$ are calculated by
\begin{eqnarray}
E[A^{est}] &=&\sum_{\{n_{i}\}}p(\{n_{i}\})A^{est}(\{n_{i}\}),\nonumber\\ 
{\rm Var}[A^{est}] &=& E[(A^{est})2]- E[A^{est}]^{2},
\end{eqnarray}
where the summation is taken over all sets $\{n_{i}\}$ that satisfy $n_{i}\geq 0$ and $\sum_{i} n_{i} = n$, and $p(\{n_{i}\})$ is the
probability that each outcome $i$ is obtained $n_{i}$ times,
\begin{eqnarray}
p(\{n_{i}\})=n!\prod_{i}\frac{p_{i}^{n_{i}}}{n_{i}!}.
\end{eqnarray} 
In our example, where $p_{i}=\sum_{l}p_{i,l}(\psi)$ in (52), the expectation value of the estimator $A^{est}$ is identified as 
\begin{eqnarray}
\lim_{n\rightarrow\infty}E[A^{est}]=\langle M^{out}\rangle=\sum_{k,l}m_{k}\langle \psi|M^{\dagger}_{k,l}M_{k,l}|\psi\rangle.
\end{eqnarray}
In our definition of $A^{est}$, we obtain
\begin{eqnarray}
\lim_{n\rightarrow\infty}n{\rm Var}(A^{est})=\sum_{i}m^{2}_{i}p_{i}-(\sum_{i}m_{i}p_{i})^{2}=\sigma(M^{out})^{2}
\end{eqnarray}
which minimizes the error arising from the statistical estimation, since $\sigma(M^{out})^{2}$ contains only the measurement errors and the intrinsic fluctuation of the initial quantum state.

They then define the "consistent estimator" by~\cite{watanabe1, watanabe2}
\begin{eqnarray}
\lim_{n\rightarrow\infty}{\rm Prob}(|A^{est}-\langle A\rangle|<\delta)=1
\end{eqnarray}
for {\em all} states and arbitrary $\delta>0$. The consistent estimator thus satisfies 
\begin{eqnarray}
\lim_{n\rightarrow\infty}E[A^{est}]=\langle M^{out}\rangle=\langle A\rangle
\end{eqnarray}
for {\it all} $\psi$, namely, the {\em consistent estimator implies the unbiased measurement} in our formulation. To generate the events for the statistical analysis one needs to measure the given system, and it may be natural to assume that the measurement is effectively described by a suitably chosen measurement operator $M$. We can then understand that the consistent estimator, which was introduced from a point of view of statistical estimation theory, provides a practical  operational definition of the unbiased measurement in the context of measurement operator formalism. 

Their definition of error~\cite{watanabe1, watanabe2}
\begin{eqnarray}
\epsilon(A; M)=\min_{A^{est}}\lim_{n\rightarrow\infty}n{\rm Var}(A^{est})-\sigma(A)^{2}
\end{eqnarray}
then agrees with the relation of the Arthurs-Kelly formulation, $\sigma(M^{out})^{2}=\epsilon(A)^{2}+\sigma(A)^{2}$ which is based on the unbiased measurement $\langle M^{out}\rangle=\langle A\rangle$, if one identifies
\begin{eqnarray}
\epsilon(A; M)=\epsilon(A)^{2}=\langle(M^{out}-A)^{2}\rangle.
\end{eqnarray}
The consistency of this identification is confirmed by the fact that the error $\epsilon(A; M)$ defined in their scheme is also non-negative and vanishes only for the precise projective measurement of $A$. 
The precise projective measurement of $A$ implies $P_{k}=\sum_{l}M_{k,l}^{\dagger}M_{k,l}$ with $P_{k}P_{k^{\prime}}=\delta_{k,k^{\prime}}P_{k}$ and the spectral decomposition of $A=\sum_{k}m_{k}P_{k}$ with $\sum_{k}P_{k}=1$.

The relation  (60) holds without assuming consistent estimators, but the consistent estimator or unbiased measurement is crucial to ensure the non-negative error $\epsilon(A; M)$ in (64) in the framework of positive operator valued measures. Since 
\begin{eqnarray}
\sigma(M^{out})^{2}-\sigma(A)^{2}&=&\sigma(M^{out}-A)^{2}\nonumber\\
&+&\langle(M^{out}-A-\langle M^{out}-A\rangle)A\rangle\nonumber\\
&+&\langle A(M^{out}-A-\langle M^{out}-A\rangle)\rangle
\end{eqnarray}
and the first term is quadratic in $M^{out}-A$ while the second and third terms are linear in $M^{out}-A$ near $M^{out}-A\sim0$ on the right-hand side,
and thus the right-hand side is indefinite for $M^{out}-A\sim0$ without the unbiased measurement condition. Conversely, if the unbiased measurement condition is satisfied, we have $\sigma(M^{out}-A)^{2}=\langle(M^{out}-A)^{2}\rangle=\langle(M^{out})^{2}\rangle-\langle (M^{out}-A)A\rangle-\langle A(M^{out}-A)\rangle-\langle A^{2}\rangle=\langle(M^{out})^{2}\rangle-\langle A^{2}\rangle=\sigma(M^{out})^{2}-\sigma(A)^{2}\geq 0$ which ensures the non-negativity of $\epsilon(A; M)$ in (64). Any sensible estimation theory is based on the well-defined error. The condition of consistent estimator is thus crucial in the present estimation theory to ensure the non-negative error (squared) defined by a difference in (63), and the entire formulation is constructed to preserve (62). At the same time, the present estimation theory works only for the unbiased measurement, while the measurement operator formalism is more flexible and  applicable to measurements without any conditions.

The unbiased measurement of $N$ is defined by
\begin{eqnarray}
\langle N^{out}\rangle=\sum_{k,l}n_{l}\langle \psi|M^{\dagger}_{k,l}M_{k,l}|\psi\rangle=
\langle \psi| B|\psi\rangle
\end{eqnarray}
for all $\psi$ or equivalently $B=\sum_{k,l}n_{l}M^{\dagger}_{k,l}M_{k,l}$, and using this unbiased condition
\begin{eqnarray}
\epsilon(B)^{2}&=&\langle(N^{out}-B)^{2}\rangle\nonumber\\
&=&\langle(N^{out})^{2}\rangle-\langle B^{2}\rangle
\nonumber\\
&=&\sigma(N^{out})^{2}-\sigma(B)^{2}\nonumber\\
&=&\sum_{k,l} n_{l}^{2}\langle \psi|M^{\dagger}_{k,l}M_{k,l}|\psi\rangle - \langle \psi|B^{2}|\psi\rangle.
\end{eqnarray}
Our analysis of (8)  suggests that the condition (66) is not satisfied by the well defined $N^{out}$ for the precise projective measurement of $A$, or the condition (66) may be satisfied by a singular $N^{out}$ which may give rise to a divergent result in (67) for the precise projective measurement of $A$.  

\subsubsection{ Heisenberg-type error-error relation}
 
From (64) and (67), we thus conclude the Heisenberg-type error-error relation
\begin{eqnarray}
\epsilon(A; M)\epsilon(B; N)\geq \frac{1}{4}|\langle [A, B]\rangle|^{2}
\end{eqnarray}
by combining Robertson's relation (12) with unbiased joint measurements, in agreement with the analysis in (19). The interesting result found by numerical simulations in~\cite{watanabe1} is that this relation is valid even for $\epsilon(A; M)\rightarrow 0$ with $\langle [A, B]\rangle\neq 0$ since $\epsilon(B; N)\rightarrow\infty$ if one strictly imposes the unbiased condition $\langle\psi| N^{out}|\psi\rangle=\langle\psi| B|\psi\rangle$ for all $\psi$; this singular behavior 
$\epsilon(B; N)\rightarrow\infty$ is not unnatural since the unbiased measurement condition even for the bounded operator $B$ may generally require quite a large unbounded measurement operator $N$. This conclusion is perfectly consistent with Theorem 4 in~\cite{ozawa2} in the measurement operator formalism if one identifies consistent estimator with unbiased measurement as we have shown, and this singular behavior is also consistent with the analysis in (8). The relation (68) also implies the validity of the standard Arthurs-Kelly relation in (20) as a conditionally valid uncertainty relation.\\

\subsubsection{Heisenberg-type error-disturbance relation}
As for the disturbance, the unbiased disturbance which satisfies $\langle B^{out}-B \rangle=0$ for all $\psi$ is not explicitly mentioned in the paper~\cite{watanabe2}. They instead introduce the optimal
measurement operator $N^{opt}$ that retrieves the maximum information about $B$
after the measurement of $A$. This is achieved in the present operator  formulation by the "precise measurement" $N^{opt}$ of $B^{out}$ defined by
\begin{eqnarray}
N^{opt}|\psi\otimes\xi\rangle=B^{out}|\psi\otimes\xi\rangle
\end{eqnarray}
for all  $\psi$, which implies 
\begin{eqnarray}
\langle\psi\otimes\xi|N^{opt}|\psi\otimes\xi\rangle
=\langle\psi\otimes\xi|B^{out}|\psi\otimes\xi\rangle
\end{eqnarray}
for all $\psi$ and also $\sigma(N^{opt})=\sigma(B^{out})$.
The relation (70) implies that $N^{opt}$ defines a consistent estimator of $B$ if one assumes the unbiased disturbance $\langle\psi\otimes\xi|B^{out}|\psi\otimes\xi\rangle=\langle\psi|B|\psi\rangle$ for all $\psi$, which in turn  implies $\sigma(B^{out})\geq\sigma(B)$.

They then define the disturbance $\eta(B; M)$ by
\begin{eqnarray}
\eta(B; M)=\sigma(N^{opt})^{2}-\sigma(B)^{2}
\end{eqnarray}
which is consistent with the unbiased disturbance $\sigma(B^{out})^{2}=\eta(B)^{2}+\sigma(B)^{2}$ if one identifies
\begin{eqnarray}
\eta(B; M)=\eta(B)^{2}=\langle (B^{out}-B)^{2} \rangle.
\end{eqnarray} 
This is consistent since the disturbance $\eta(B; M)$ is non-negative and vanishes for the disturbance-free case $B^{out}=B$. 
In general we have no control of the distribution of $B$ after the measurement of $A$, thus the implementation of the unbiased  disturbance is more subtle than the unbiased measurement. In view of (54) we need to satisfy 
\begin{eqnarray}
\langle B^{out}\rangle=\sum_{k,l}\langle \psi|M^{\dagger}_{k,l}BM_{k,l}|\psi\rangle=\langle \psi|B|\psi\rangle
\end{eqnarray}
for all $\psi$, or $B=\sum_{k,l}M^{\dagger}_{k,l}BM_{k,l}$.

The error-disturbance relation proposed on the basis of the above analysis in~\cite{watanabe2}
\begin{eqnarray}
\epsilon(A; M)\eta(B; M)\geq \frac{1}{4}|\langle [A, B]\rangle|^{2}
\end{eqnarray}
{\em formally} agrees with the naive Heisenberg-type error-disturbance relation (37) proposed by Ozawa~\cite{ozawa1}, which is derived from Robertson's relation (28) by assuming unbiased measurement and disturbance, as was emphasized in~\cite{fujikawa3}. 

Two different interpretations of (74) are possible: The first one adopted in~\cite{ozawa1,hasegawa} is to identify
\begin{eqnarray}
\eta(B; M)=\langle (B^{out}-B)^{2} \rangle\leq ||(B^{out}-B)^{2} ||\leq 4||B||^{2}.
\end{eqnarray}
In this interpretation, (74) is bound to fail for $\epsilon(A; M)\rightarrow 0$ for the bounded $||B||<\infty$ and $\langle [A, B]\rangle\neq 0$, as was demonstrated in~\cite{hasegawa}. This failure of (74) arises from the failure of its derivation since the crucial assumption of the unbiased disturbance fails for $\epsilon(A; M)\rightarrow 0$ as is indicated by (11).

The other interpretation which is adopted in~\cite{watanabe2} is to identify
\begin{eqnarray}
\eta(B; M)=\langle (N^{opt})^{2} \rangle-\langle B^{2} \rangle=\sigma(N^{opt})^{2}-\sigma(B)^{2}.
\end{eqnarray}
In this interpretation, (74) is essentially the same as (68) and it holds even for $\epsilon(A; M)\rightarrow 0$ with $\langle [A, B]\rangle\neq 0$ by formally letting $\sigma(N^{out})\rightarrow\infty$. But a consequence of the precise measurement  condition  $\sigma(N^{out})=\sigma(B^{out})<\infty$ for the bounded operator is lost, namely, the condition of consistent estimation fails.

We note that the quantity  $\eta(B; M)$ in (72) and (74) is finite for the {\em bounded}  operator $B$ in any sensible definition of "disturbance". For example,  $\langle (B^{out})^{2} \rangle$ on the right-hand side of (54) which is based on positive operator valued measures gives  
\begin{eqnarray}
\sum_{k}\langle \psi|M^{\dagger}_{k}B^{2}M_{k}|\psi\rangle
&\leq& \sum_{k}|\langle \psi|M^{\dagger}_{k}B^{2}M_{k}|\psi\rangle|\nonumber\\
&=& \sum_{k^{\prime}}|\frac{\langle \psi|M^{\dagger}_{k^{\prime}}B^{2}M_{k^{\prime}}|\psi\rangle}{\langle \psi|M^{\dagger}_{k^{\prime}}M_{k^{\prime}}|\psi\rangle}|\langle \psi|M^{\dagger}_{k^{\prime}}M_{k^{\prime}}|\psi\rangle\nonumber\\
&\leq&\sum_{k^{\prime}}||B^{2}||\langle \psi|M^{\dagger}_{k^{\prime}}M_{k^{\prime}}|\psi\rangle\nonumber\\
&=&||B^{2}||\leq ||B||^{2}
\end{eqnarray}
where the summation over $k^{\prime}$ means that the terms with $\langle \psi|M^{\dagger}_{k}M_{k}|\psi\rangle=0$ are excluded since $\langle \psi|M^{\dagger}_{k}M_{k}|\psi\rangle=0$ implies $M_{k}|\psi\rangle=0$. The quantity 
$\langle (B^{out})^{2} \rangle$, which is less than $||B||^{2}$ for the well-defined measurement, cannot go to $\infty$ in a discontinuous manner.
\\

 We identify the essence of quantum estimation theory~\cite{watanabe1,watanabe2} with the assumptions of {\em consistent} (or unbiased) measurement and disturbance when it is applied to the analysis of uncertainty relations. The numerical simulation in quantum estimation theory~\cite{watanabe1} shows that the error-error uncertainty relation (68) together with the Arthurs-Kelly relation (20) and (23) are maintained, although these relations are conditionally valid depending on strict unbiased conditions. On the other hand, we prefer the interpretation of the error-disturbance relation (74) in the manner of references~\cite{ozawa1, hasegawa}, namely, the relation (74) {\em fails} by preserving the finiteness of the disturbance of the bounded operators such as spin variables.  Of course, the failure of (74) does not imply the failure of the "Heisenberg uncertainty relation" as such but rather it implies that the derivation of (74) from the universally valid Robertson's relation fails just as the failure of the relation (46) in the spin measurement. The universally valid Heisenberg relation (33) always holds.

\section{Discussion and conclusion}
We emphasized the algebraic incompatibility of the precise measurement of one of the conjugate variables with the assumptions of unbiased joint measurements or unbiased measurement and disturbance if all the operators involved are well-defined, independently of uncertainty relations. 

We clarified the origin of the different conclusions concerning conditionally valid uncertainty relations in the existing literature by pointing out that consistent estimator in estimation theory~\cite{watanabe1,watanabe2} corresponds to unbiased measurement and disturbance.
The consistent quantum estimation is possible for the Heisenberg-type error-error relation (68) and the standard Arthurs-Kelly relation (20) by allowing the singular behavior of measurement operators. The consistency of the Arthurs-Kelly relation, although it is  valid only conditionally on the basis of unbiased joint measurements, is important for practical applications~\cite{she, yuen, yamamoto}. On the other hand, we argued that the consistent quantum estimation fails for the naive Heisenberg-type error-disturbance relation (74) for bounded operators~\cite{ozawa1} since it requires the divergent disturbance of bounded operators, contrary to the physical picture of disturbance.

In contrast, the universally valid uncertainty relations such as (23), (33) and (36), which are formulated without imposing any extra conditions, are always valid.  
 As for the candidate of the original Heisenberg uncertainty relation, we prefer the universally valid Heisenberg relation (33) by adopting the small "inaccuracy" $\bar{\epsilon}(A)=\epsilon(A)+\sigma(A)$, namely, a precise measurement with small $\epsilon(A)$ of a well-defined state with small $\sigma(A)$,  as a criterion of a "good measurement"  (or simply the universally valid Robertson's relation (36) itself if you ask the saturation of the inequality).

As for the analysis of unbounded operators such as $\hat{p}$ and $\hat{x}$, the mathematics involved is subtle but 
we  expect that our conclusion concerning the naive Heisenberg-type error-disturbance relation, namely, its failure  still holds although our argument on the basis of bounded operator is not applicable. One way to see this may be to start with a "regularized" expression such as mentioned in Footnote 2.  Besides, the practical implementation of the unbiased disturbance is expected to be difficult, and thus the naive Heisenberg-type error-disturbance relation is not derived from Robertson's relation. (The unbiased disturbance may be implemented by means of "selective sampling" of events~\cite{fujikawa4} for which the behavior of the uncertainty relation can be unconventional.)  In contrast, the universally valid versions of uncertainty relations are expected to be always valid for unbounded hermitian operators also if properly formulated, and the universally valid relation such as (33) or (36) implies that $\eta(B)$ can stay finite for $\epsilon(A)=0$ if one does not impose the unbiased disturbance condition.
\\

I am grateful to S. Tanimura for calling the works~\cite{watanabe1, watanabe2}
to my attention. I also thank M. Ueda for very helpful and clarifying comments on quantum estimation theory.

\end{document}